\RequirePackage{fix-cm}
\documentclass[twocolumn,epjc3]{svjour3}  
\smartqed  % flush right qed marks, e.g. at end of proof
\RequirePackage{epsfig}
\RequirePackage{url}
\RequirePackage{hyperref}

\RequirePackage[normalem]{ulem} 

\RequirePackage{latexsym}
\RequirePackage{epsfig}
\RequirePackage{amsmath}
\RequirePackage{amssymb}
\RequirePackage{wasysym}
\RequirePackage{graphicx}
\RequirePackage{verbatim}
\RequirePackage{enumerate,mdwlist}%lists
\RequirePackage[titletoc]{appendix}
\RequirePackage{amsfonts}
\RequirePackage{tikz} %diagrams
\usetikzlibrary{calc}
\RequirePackage{pgfplots}
\RequirePackage[export]{adjustbox}

\newcommand{\HCd}{\mathcal{H}}

\def\HCdt0{\tilde{\HCd}_{0}}

\newcommand\rf[1]{(\ref{eq:#1})}
\newcommand\lab[1]{\label{eq:#1}}
\newcommand\nonu{\nonumber}
\newcommand\br{\begin{eqnarray}}
\newcommand\er{\end{eqnarray}}
\newcommand\be{\begin{equation}}
\newcommand\ee{\end{equation}}

\newcommand\lb{\lbrack}
\newcommand\rb{\rbrack}

\renewcommand\){\right)}
\newcommand\bgv{\bigg\vert}              %%

\newcommand\bc{\begin{center}}
\newcommand\ec{\end{center}}

\newcommand\partder[2]{\frac{{\partial {#1}}}{{\partial {#2}}}}
\renewcommand\d{\delta}

\newcommand\vareps{\varepsilon}

\newcommand\G{\Gamma}

\newcommand\h{\frac{1}{2}}
\renewcommand\k{\kappa}
\renewcommand\l{\lambda}
\renewcommand\L{\Lambda}
\newcommand\m{\mu}
\newcommand\n{\nu}
\newcommand\om{\omega}

\newcommand\vp{\varphi}
\renewcommand\P{\Phi}
\newcommand\pa{\partial}

\newcommand\pr{\prime}

\newcommand\z{\zeta}

\newcommand\twomat[4]{\left(\begD{array}{cc}  %%   2x2 matrix  %ESA
{#1} & {#2} \\ {#3} & {#4} \end{array} \right)}

\newcommand\cM{{\mathcal M}}
\newcommand\cN{{\mathcal N}}

\newcommand{\ct}[1]{\cite{#1}}
\newcommand\PRD[3]{{Phys. Rev.} \textbf{D#1}, #3 (#2)}

\newcommand\PLB[3]{{Phys. Lett.} \textbf{#1B}, #3 (#2)}
\newcommand\CQG[3]{{Class. Quantum Grav.} \textbf{#1}, #3 (#2)}

\newcommand\PRep[3]{{Phys. Reports} \textbf{#1}, #3 (#2)}

\newcommand\IJMPA[3]{{Int. J. Mod. Phys.} \textbf{A#1}, #3 (#2)}
\newcommand\IJMPD[3]{{Int. J. Mod. Phys.} \textbf{D#1}, #3 (#2)}

\newcommand\udot{\stackrel{.}{u}}
\newcommand\uddot{\stackrel{..}{u}}

\newcommand\adot{\stackrel{.}{a}}

\newcommand\Hdot{\stackrel{.}{H}}

\RequirePackage{mathptmx}      % use Times fonts if available on your TeX system
\RequirePackage{latexsym}
\RequirePackage[numbers,sort&compress]{natbib}
\journalname{}
\begin{document}
\title{Dynamically Generated Inflation from Non-Riemannian Volume Forms} 


\author{D. Benisty\thanksref{e1,addr1,addr2}
        \and
E. I. Guendelman\thanksref{e2,addr1,addr2,addr3} %etc.
\and E. Nissimov \thanksref{e3,addr4}
       \and S. Pacheva \thanksref{e4,addr4}
}

\thankstext{e1}{e-mail: benidav@post.bgu.ac.il}
\thankstext{e2}{e-mail: guendel@bgu.ac.il}
\thankstext{e3}{e-mail: nissimov@inrne.bas.bg}
\thankstext{e4}{e-mail: svetlana@inrne.bas.bg}

\institute{Physics Department, Ben-Gurion University of the Negev, Beer-Sheva 84105, Israel \label{addr1}
\and
Frankfurt Institute for Advanced Studies (FIAS), Ruth-Moufang-Strasse~1, 60438 Frankfurt am Main, Germany \label{addr2}
\and
Bahamas Advanced Study Institute and Conferences, 4A Ocean Heights, Hill View Circle, Stella Maris, Long Island, The Bahamas  \label{addr3}
           \and
Institute for Nuclear Research and Nuclear Energy, Bulgarian Academy of Sciences, Sofia, Bulgaria \label{addr4}
}

\date{}

\maketitle

\begin{abstract}
We propose a simple modified gravity model {\em without} any initial matter
fields in terms of several alternative non-Rie\-mann\-ian
spacetime volume elements within the metric (second order) formalism. We show how the 
non-Rie\-mann\-ian volume-elements,
when passing to the physical Einstein frame, create a canonical scalar
field and produce dynamically a non-trivial inflationary-type potential for
the latter with a large flat region and a stable low-lying minimum.
We study the evolution of the cosmological solutions from the point of view
of theory of dynamical systems. The theory predicts the spectral index $n_s \approx 0.96$ and the tensor-to-scalar ratio $r \approx 0.002$ for 60 $e$-folds, which is in accordance with the
observational data. In the future Euclid and SPHEREx missions or the BICEP3 experiment are expected to provide experimental evidence to test those predictions.
\end{abstract}

%%%%%% change-begin %%%%%%
\section{Introduction}
%%%%%% change-end %%%%%%

Developments in cosmology have been influenced to a great extent by the idea of inflation 
\cite{Starobinsky:1979ty,Starobinsky:1980te,Linde:1981mu,Guth:1980zm,Albrecht:1982wi}, which provides an attractive solution
of the fundamental puzzles for the standard Big Bang model, as the horizon and the flatness problems. 
In addition, providing a framework for sensible calculations of primordial density perturbations were 
discussed in \cite{Mukhanov:1981xt,Guth:1982ec}. However, it has been recognized that a successful
implementation requires some very special restrictions on the dynamics that drives inflation. 
In particular, in {\em New Inflation} \ct{Linde:1981mu}, a potential with a large flat region, which then 
drops to zero (or almost zero) in order to reproduce the vacuum with almost zero (in Planck units) 
cosmological constant of the present universe, is required.

In a parallel development, extended (modified) gravity theories as 
alternatives/generalizations of the standard Einstein General Relativity are 
being extensively studied in the last decade or so. 
%%%%%% chenge-begin %%%%%%
The main motivation for this development comes from:

\begin{itemize}
\item
(a) Cosmology -- modified gravity may solve the problems of dark energy 
and dark matter and explain the large scale structure and the accelerated
expansion of the universe \cite{Perlmutter:1998np,Copeland:2006wr}); 
\item
(b) Quantum field theory in curved spacetime -- because of non-renormalizability 
of standard general relativity in higher loops it fails to describe the
universe at quantum scales \ct{weinberg-79}; 
\item
(c) Modern String theory -- because of the natural appearance of
scalar-tensor couplings and higher-order
curvature invariants in low-energy effective field theories aimed at
phenomenologically realistic description of particle physics \ct{GSW-1}. 
\end{itemize}

The principal approaches to construct modified gravity theories include 
$f(R)$-gravity, scalar-tensor theories, Gauss-Bonnet gravity models. 
For detailed accounts, see the book \ct{extended-grav-book} and the
extensive reviews \ct{extended-grav,odintsov-1,berti-etal,odintsov-2}, as well as for
further details Refs.\cite{Dvali:1998pa}-\cite{Benisty:2019jqz}. 
%\cite{Dvali:1998pa,Kawasaki:2000yn,Bojowald:2002nz,Nojiri:2003ft,Kachru:2003sx,Nojiri:2005pu,Ferraro:2006jd,Cognola:2007zu,Cai:2010kp,Ashtekar:2011rm,Qiu:2011zr,Briscese:2012ys,Ellis:2013xoa,Basilakos:2013xpa,Sebastiani:2013eqa,Baumann:2014nda,Dalianis:2015fpa,Kanti:2015pda,DeLaurentis:2015fea,Basilakos:2015yoa,Bonanno:2015fga,Koshelev:2016xqb,Bamba:2016wjm,Motohashi:2017vdc,Oikonomou:2017ppp,Benisty:2018fja,Antoniadis:2018ywb,Karam:2019dlv,Nojiri:2019kkp,Benisty:2019jqz}.
%. 
%%%%% change-end %%%%%%

%% 4D-GB2
One broad class of actively developed modified/extended gravitational theories is 
based on employing alternative non-Riemannian spacetime volume-forms, 
\textsl{i.e.}, metric- independent generally 
covariant volume elements in the pertinent Lagrangian actions
%%%% change-begin %%%%%%
on spacetime manifolds with an ordinary Riemannian geometry,
instead of the canonical Riemannian volume element $\sqrt{-g} \, d^4 x$
whose density is given by the square-root of the determinant of the Riemannian metric:
\begin{equation}
 \sqrt{-g} \equiv \sqrt{-\det\Vert g_{\m\n}\Vert}  
\end{equation}
originally proposed in  \ct{TMT-orig-0,Hehl,TMT-orig-1,TMT-orig-2,5thforce}.
For a concise geometric formulation, see \ct{susyssb-1,grav-bags}.
%%%% change-end %%%%%%

This formalism was used as a basis for constructing a series of extended gravity-matter
models describing unified dark energy and dark matter scenario \ct{dusty,dusty-2},
quintessential cosmological models with gravity-assisted and inflaton-assisted
dynamical suppression (in the ``early'' universe) or generation (in the
post-inflationary universe) of electroweak spontaneous symmetry
breaking and charge confinement \ct{grf-essay,varna-17,bpu-10}, and a novel 
mechanism for the supersymmetric Brout-Englert-Higgs effect in supergravity 
\ct{susyssb-1}.

In the present paper we propose a very simple gravity model {\em without} any
initial matter fields involving several
non-Riemannian volume-forms instead of the standard Riemannian 
volume element $\sqrt{-g} \, d^4 x$. We show how the non-Riemannian volume-elements, when passing to the physical Einstein frame, generate a canonical scalar field $u$ 
and manage to create dynamically a non-trivial inflationary-type potential for $u$ 
with a large flat region for large positive $u$ and a stable low-lying minimum,
\textsl{i.e.}, $u$ will play the role of a {\em dynamically created} ``inflaton''.
%%%%% change-begin %%%%%%
This dynamically generated inflationary potential turns out to be a generalization 
of the well-known Starobinsky potential \ct{Starobinsky:1979ty}.
%%%%% change-begin %%%%%%

We study the evolution of the cosmological solutions from the point of view
of the theory of dynamical systems and calculate the spectral index $n_s$
and the tensor-to-scalar ratio $r$ in our model whose values are in accordance 
with the observational data.

%% NOTE ADDED IN PROOF
In Section 2 below we briefly review the general notion of volume forms on arbitrary differential manifolds. Section 3 briefly presents the general construction of Lagrangian actions on Riemannian manifolds employing metric-independ\-ent (non-Riemannian) volume forms (volume elements). Our main results are contained in Sections 4, 5 and 6. In Section 4 we propose our simple modified gravity model in terms of several non-Riemannian volume elements without any matter fields and derive the corresponding Einstein-frame description with the associated dynamical creation of a canonical scalar field with a non-trivial effective inflationary potential. In Section 5 we study the cosmological evolutionary solutions within the Friedmann-Lemaitre-Robertson-Walker framework. In Section 6 we derive the explicit expressions for the Hubble slow-roll parameters and use them to obtain analytic results for the scalar power spectral index and the tensor-to-scalar ratio which we compare with the available observational  data.The last Section 7 contains our conclusions.
%%%%%%%%%%%%%%%%%%%%%%%%%%%%%%%%%%%%%%%%%%%%%%%%%%%%%%%%%%%%%%%%%%%%%%%%%%%%%%%%%%%%%%%%%%%%%%%%
%%%%%%%%%%%%%%%%%%%%%%%%%%%%%%%%%%%%%%%%%%%%%%%%%%%%%%%%%%%%%%%%%%%%%%%%%%%%%%%%%%%%%%%%%%%%%%%%
\section{Non-Riemannian Volume-Forms Formalism}

%%%%% change-begin %%%%%%
Let us first recall the general notion of  volume-forms (volume elements) 
in integrals over arbitrary differentiable manifolds -- not necessarily Riemannian one,
so {\em no} metric is needed. Volume forms are given by nonsingular maximal rank 
differential forms $\om$ (see e.g. Ref.\ct{spivak}):
%%%%% change-end %%%%%%
\br
\int_{\cM} \om \bigl(\ldots\bigr) = \int_{\cM} dx^D\, \Omega \bigl(\ldots\bigr)
\;\; ,
\nonu \\
\om = \frac{1}{D!}\om_{\m_1 \ldots \m_D} dx^{\m_1}\wedge \ldots \wedge dx^{\m_D}\; ,
\lab{omega-1} \\
\om_{\m_1 \ldots \m_D} = - \vareps_{\m_1 \ldots \m_D} \Omega \; ,
% \;\; ,\;\;
% dx^{\m_1}\wedge \ldots \wedge dx^{\m_D} = \vareps^{\m_1 \ldots \m_D}\,  dx^D \; ,
% \lab{omega-3}
\nonu
\er
(our conventions for the alternating symbols $\vareps^{\m_1,\ldots,\m_D}$ and
$\vareps_{\m_1,\ldots,\m_D}$ are: $\vareps^{01\ldots D-1}=1$ and
$\vareps_{01\ldots D-1}=-1$).

%%%%%% change-begin %%%%%%
The volume element density $\Omega$, as it is evident from its definition in
\rf{omega-1}, transforms as scalar density under general coordinate 
reparametrizations on the manifold.
%%%%%% change-end %%%%%%

In Riemannian $D$-dimensional spacetime manifolds a standard generally-covariant 
volume-form is defined through the ``D-bein'' (frame-bundle) canonical one-forms 
$e^A = e^A_\m dx^\m$ ($A=0,\ldots ,D-1$):
\br
\om = e^0 \wedge \ldots \wedge e^{D-1} = \det\Vert e^A_\m \Vert\,
dx^{\m_1}\wedge \ldots \wedge dx^{\m_D} % \quad  
\nonu \\
\longrightarrow \quad
\Omega = \det\Vert e^A_\m \Vert = \sqrt{-\det\Vert g_{\m\n}\Vert} \; .
\lab{omega-riemannian}
\er

To construct modified gravitational theories as alternatives to ordinary
standard theories in Einstein's general relativity, instead of $\sqrt{-g}$ 
we can employ one or more alternative {\em non-Riemannian} 
volume element densities
as in \rf{omega-1} given by non-singular {\em exact} $D$-forms
$\om = d A$ where:
\br
A = \frac{1}{(D-1)!} A_{\m_1\ldots\m_{D-1}}
dx^{\m_1}\wedge\ldots\wedge dx^{\m_{-1}} % \quad \longrightarrow \quad
\nonu \\
% \lab{B-form}
% \ee
% so that the {\em non-Riemannian} volume element reads:
% \be
\longrightarrow \quad  \Omega \equiv \Phi(A) =
\frac{1}{(D-1)!}\vareps^{\m_1\ldots\m_D}\, \pa_{\m_1} A_{\m_2\ldots\m_D} \; .
\lab{Phi-D}
\er
Thus, the non-Riemannian volume element density $\Phi (A)$ is defined in terms of
the (scalar density of the) dual field-strength of an auxiliary rank 
$D-1$ tensor gauge field $A_{\m_1\ldots\m_{D-1}}$ 
%%%%% change-begin %%%%%%
and it transforms as scalar density under general coordinate transformations,
which is evident from its definition \rf{Phi-D}. Accordingly, the integration
element $\int d^4 x \Phi (A)$ is manifestly invariant under general
coordinate transformations.

Let us stress that the term ``non-Riemannian'' relates only to the nature of the
volume element density \rf{Phi-D}, whose definition does not involve the metric.
Otherwise the geometry of the spacetime is a regular Riemannian one --
scalar products of vector fields are given as usual by the Riemannian
metric $g_{\m\n}$, the connection $\G_{\m\n}^\l$ is the usual Levi-Civita one in terms of $g_{\m\n}$,
there is {\em no} torsion, \textsl{etc.}.
%%%%% change-end %%%%%%
\section{The Action}
In general, modified gravity Lagrangian actions based on the non-Riemannian
volume-form formalism have the following generic form (here and below we are
using units with $16 \pi G_{\rm Newton} = 1$):
\be
S = \int d^4 x \Bigl\{ \P_1 (A) \Bigl\lb R + L^{(1)}\Bigr\rb
+ \P_2 (B)\Bigl\lb L^{(2)} + \frac{\P_0 (C)}{\sqrt{-g}}\Bigr\rb + \ldots\Bigr\} \;.
\lab{NRVF-0}
\ee
Here $\P_1 (A), \P_2 (B), \P_0 (C)$ are several different non-Rie\-mann\-ian volume
element densities of the form \rf{Phi-D}, \textsl{i.e.}, defined by auxiliary 
rank 3 tensor gauge fields $A_{\m\n\l} , B_{\m\n\l} , C_{\m\n\l}$; 
$R$ denotes the scalar curvature in either first-order
(Palatini) or second order (metric) formalism; $L^{(1)}$ and $L^{(2)}$ are
some matter field Lagrangians; 
% $\P_0 (C)$ is given by a
%third auxiliary tensor gauge field $C_{\m\n\l}$ as in \rf{Phi-D} and is needed for
%consistency of the formalism; 
the dots indicate possible additional terms
containing higher powers of the non-Riemannian volume element densities
\textsl{e.g.}, $\bigl(\P_1 (A)\bigr)^2/\sqrt{-g}$. % or $\P_2 (B)\P_0(C)/\sqrt{-g}$. 
The specific forms of $L^{(1)}$ and $L^{(2)}$
can be uniquely fixed via the requirement for invariance of \rf{NRVF-0}
under global Weyl-scale invariance (see \rf{scale-transf} below).

%%%%% change-begin %%%%%%
Let us stress that the modified gravity action \rf{NRVF-0}, in complete
analogy with 
\begin{equation}
\int d^4 x\, \sqrt{-g}\Bigl\lb R + \ldots\Bigr\rb
\end{equation}
which is the standard Ein\-stein-Hilbert action, is explicitly invariant under general coordinate reparametrizations since, as mentioned above, non-Rie\-mann\-ian volume element densities transform as scalar densities similarly to $\sqrt{-g}$.
%%%%% change-end %%%%%%

A characteristic feature of the modified gravitational theories \rf{NRVF-0} is that
when starting in the first-order (Palatini) formalism all non-Riemannian
volume-forms are almost {\em pure-gauge} degrees of freedom, \textsl{i.e.} 
they {\em do not} introduce any additional 
%%%%% change-begin %%%%%%
physical (field-propagating) gravitational degrees of freedom 
except for few discrete degrees 
of freedom with conserved canonical momenta appearing as arbitrary integration 
constants. The reason is that the modified gravity action in Palatini
formalism is linear w.r.t. the velocities of some of the auxiliary gauge
field components defining the non-Riemannian volume element densities, 
and does not depend on the velocities of the rest of auxiliary gauge field
components.
The (almost) pure-gauge nature of the latter is explicitly shown 
in Refs.\ct{grav-bags,grf-essay} (appendices A) employing the standard 
canonical Hamiltonian treatment of systems with gauge symmetries, i.e.,
systems with first-class Hamiltonian constraints a'la Dirac (\textsl{e.g.},
\ct{henneaux-teitelboim,rothe}).

Unlike Palatini formalism, the above situation changes significantly
when we treat \rf{NRVF-0} in the second order (metric) formalism. In the
latter case the ``Einstein-Hilbert'' part $\int d^4 x\, \P_1 (A) R$
of the modified gravity action \rf{NRVF-0} contains sec\-ond order time
derivative terms of the metric in $R$, which is in sharp contrast with the
case of ordinary Riemannian volume element $\int d^4 x\, \sqrt{-g} R$
where the corresponding second-order time derivatives amount to a total derivative.
According to the general canonical
Hamiltonian treatment of systems with higher-order time derivatives on the
canonical variables (see \textsl{e.g.}, \ct{gitman-tyutin} -- modern version
of the classical Ostrogradsky formalism \ct{ostrogradski}) the presence of the latter 
implies the appearance of some of the corresponding velocities as additional 
physical degrees of freedom. In the present case this is reflected in the fact 
that (as we will see below, Eqs.\rf{U-eff}-\rf{EF-action}) upon
%%%%% change-end %%%%%%
passing to the physical Einstein frame via
conformal transformation:
\be
g_{\m\n} \to {\bar g}_{\m\n}= \chi_1 g_{\m\n} \quad ,\quad
\chi_1 \equiv \frac{\P_1 (A)}{\sqrt{-g}} \; ,
\lab{g-bar}
\ee
the first non-Riemannian volume element density
$\P_1 (A)$ in \rf{NRVF-0} is not any more 
a ``pure gauge'', but creates a new dynamical canonical scalar field 
$u$ via $\chi_1 = \exp{\frac{u}{\sqrt{3}}}$. In the following Section we will
see how a non-trivial inflationary potential for $u$ is dynamically generated.

%%%%%%%%%%%%%%%%%%%%%%%%%%%%%%%%%%%%%%%%%%%%%%%%%%%%%%%%%%%%%%%%%%%%%%%%%%%%%%%%%%%%%%%%%%
%%%%%%%%%%%%%%%%%%%%%%%%%%%%%%%%%%%%%%%%%%%%%%%%%%%%%%%%%%%%%%%%%%%%%%%%%%%%%%%%%%%%%%%%%%
\section{Einstein Frame - the Effective Scalar Potential}

Let us now consider the simplest member in the class of modified
gravitational models \rf{NRVF-0} with {\em no} original matter fields.
\textsl{i.e.}, $L^{(1)}=0$ and  $L^{(2)}=0$, and where we only
add a quadratic term w.r.t. non-Riemannian volume element density $\P_1 (A)$:
\be
S = \int d^4 x \Bigl\{\P_1 (A)\Bigl\lb R % -\h g^{\m\n} \pa_\m\vp \pa_\n\vp 
- 2\L_0 \frac{\P_1 (A)}{\sqrt{-g}}\Bigr\rb 
% \nonu \\
+ \P_2 (B) \frac{\P_0 (C)}{\sqrt{-g}}\Bigr\} \; , %\phantom{aaaaaa}
\lab{NRVF-1}
\ee
Here $R$ is the scalar curvature in the second order (metric) formalism and:
\br
\P_1 (A) \equiv \frac{1}{3!}\vareps^{\m\n\k\l} \pa_\m A_{\n\k\l} \;,\;
\P_2 (B) \equiv \frac{1}{3!}\vareps^{\m\n\k\l} \pa_\m B_{\n\k\l} \;,
\nonu \\
\P_0 (C) \equiv \frac{1}{3!}\vareps^{\m\n\k\l} \pa_\m C_{\n\k\l} \; .
\phantom{aaaaaa}
\lab{Phi-D4}
\er
The specific form of the action \rf{NRVF-1} is dictated by the requirement
about global Weyl-scale invariance under:
\br
g_{\m\n} \to \l g_{\m\n} \;,\phantom{aaaaaaaaaa}% \G^\m_{\n\l} \to \G^\m_{\n\l} \; ,\; 
% \vp \to \vp \; ,
\nonu \\
A_{\m\n\k} \to \l A_{\m\n\k} \; ,\; B_{\m\n\k} \to \l^2 B_{\m\n\k} \; ,\; 
C_{\m\n\k} \to C_{\m\n\k} \; .
\lab{scale-transf}
\er
where $\lambda = \text{const}$. 

Scale invariance has always played an important role since the original papers on the
non-canonical volume-form formalism \ct{TMT-orig-1}.

%%%%%% change-begin %%%%%%
In a more general context let us recall, that scale invariance is a symmetry 
which relates small scales to large scales. As such 
(together with conformal symmetry) it plays fundamental role in quantum field 
theory and modern string theory in particle physics at (ultra)high energies 
as it dynamically generates (via spontaneous breakdown) mass scales hierarchies. 
On the other hand it plays an important role in cosmology as well, where it leads 
naturally to flat inflationary potentials (in the present context this is 
because it introduces a shift symmetry of the scalar field(s)) and produces candidates 
for dark matter (see the lectures at CERN's Workshop on 
{\em ``Scale invariance in particle physics and cosmology''}, Ref.\ct{CERN-workshop}).
Also let us note another specific application of spontaneously broken dilatation 
symmetry in combination with application of the non-canonical volume form
formalism: elimination of the Fifth Force Problem in a quintessential inflationary 
scenario \ct{5thforce}.
%%%%%% change-end %%%%%%

The equations of motion resulting from \rf{NRVF-1} upon variation w.r.t. 
the auxiliary gauge fields 
$A_{\m\n\l},\, B_{\m\n\l}\,, C_{\m\n\l}$ yield, respectively:
\br
R % -\h g^{\m\n} \pa_\m \vp \pa_\n \vp 
- 4\L_0 \frac{\P_1 (A)}{\sqrt{-g}} = - M_1 \equiv {\rm const} \; ,
\lab{A-eq} \\
\frac{\P_0 (C)}{\sqrt{-g}} = - M_2 \equiv {\rm const} \;\; , \;\; 
\frac{\P_2 (B)}{\sqrt{-g}} = \chi_2 \equiv {\rm const} \; .
\lab{C-B-eq}
\er
Here $M_1 , M_2$ and  $\chi_2$ are (dimensionful and dimensionless,
respectively) integration constants. The appearance of $M_1 , M_2$ indicate
spontaneous breaking of global Weyl symmetry \rf{scale-transf}.

The equations of motion w.r.t. $g_{\m\n}$ from \rf{NRVF-1} read:
%%%%%% change-begin %%%%%%
\be
R_{\m\n} - \L_0\chi_1\, g_{\m\n}
+ \frac{1}{\chi_1}\bigl( g_{\m\n} \Box{\chi_1} - \nabla_\m \nabla_\n \chi_1\Bigr)
- \frac{\chi_2 M_2}{\chi_1} g_{\m\n} = 0 \; ,
\lab{einstein-like}
\ee
with $\chi_1$ as in \rf{g-bar}. 
On the other hand, taking the trace of \rf{einstein-like} and using
Eq.\rf{A-eq} we obtain the equation of motion for $\chi_1$:
\be
3 \frac{\Box \chi_1}{\chi_1} - \frac{4\chi_2 M_2}{\chi_1} - M_1 = 0 \; .
\lab{chi1-eq}
\ee

We now transform Eqs.\rf{einstein-like} and \rf{chi1-eq} via the
conformal transformation \rf{g-bar} and show that the transformed equations
acquire the standard form of Einstein equations w.r.t. the 
new ``Einstein-frame'' metric ${\bar g}_{\m\n}$. To this end we are using
the known formulas for the conformal transformations of $R_{\m\n}$ and
$\Box\Psi$, the latter being an arbitrary scalar field, in particular 
$\Psi \equiv \chi_1$ (see \textsl{e.g.} Ref.\ct{dabrowski}; bars indicate 
magnitudes in the ${\bar g}_{\m\n}$-frame):
\begin{equation}
R_{\m\n}(g) = R_{\m\n}(\bar{g}) - 3 \frac{{\bar g}_{\m\n}}{\chi_1}
{\bar g}^{\k\l} \pa_\k \chi_1^{1/2} \pa_\l \chi_1^{1/2} 
\lab{dabrowski-1}  
\end{equation}
\begin{equation*}
\quad+ \chi_1^{-1/2}\bigl({\bar \nabla}_\m {\bar \nabla}_\n \chi_1^{1/2} +
{\bar g}_{\m\n} {\bar{\Box}}\chi_1^{1/2}\bigr) \; ,
\end{equation*}
and
\begin{equation}
\Box \chi_1 = \chi_1 \Bigl({\bar{\Box}}\chi_1 
- 2{\bar g}^{\m\n} \frac{\pa_\m \chi_1^{1/2} \pa_\n \chi_1}{\chi_1^{1/2}}\Bigr)
\; ,
\lab{dabrowski-2}    
\end{equation}
Following the analogous derivation in Ref.\ct{benisty-1}, upon using
\rf{dabrowski-1}-\rf{dabrowski-2} we rewite Eqs.\rf{einstein-like} as:
%%%%%% change-end %%%%%%
\br
R_{\m\n}(\bar{g}) - \h {\bar g}_{\m\n} R(\bar{g}) =
\nonu \\
\h \Bigl\lb % \h \pa_\m \vp \pa_\n \vp + 
\pa_\m u \pa_\n u
% \nonu \\
- {\bar g}_{\m\n}\bigl(\h {\bar g}^{\k\l} \pa_\k u \pa_\l u
%(\pa_\k\vp \pa_\l\vp + \pa_\k u \pa_\l u)
+ U_{\rm eff} (u)\bigr)\Bigr\rb \; ,
\lab{EF-eqs}
\er
where we have redefined:
\be
\P_1 (A)/\sqrt{-g}\equiv \chi_1 = \exp{\bigl(u/\sqrt{3}\bigr)}
\lab{u-def}
\ee
in order to obtain a canonically normalized kinetic term for the scalar
field $u$, and where:
\be
U_{\rm eff} (u) = 2 \L_0 - M_1 \exp{\bigl(-\frac{u}{\sqrt{3}}\bigr)} 
+ \chi_2 M_2 \exp{\bigl(-2 \frac{u}{\sqrt{3}}\bigr)} \; .
\lab{U-eff}
\ee
%%%%%% change-begin %%%%%%
On the other hand, using \rf{dabrowski-2} we rewrite Eq.\rf{chi1-eq} in
terms of the canonical scalar field $u$:
\be
{\bar{\Box}} u + \partder{U_{\rm eff}}{u} = 0 
\lab{u-eq-orig}
\ee
with $U_{\rm eff}$ as in \rf{U-eff}.
%%%%%% change-end %%%%%%

Accordingly, the corresponding Einstein-frame action reads:
\be
S_{\rm EF} = \int d^4 x \sqrt{-{\bar g}} \Bigl\lb R({\bar g}) 
% - \h {\bar g}^{\m\n}\pa_\m \vp \pa_\n \vp 
% \nonu\\
- \h {\bar g}^{\m\n}\pa_\m u \pa_\n u  
- U_{\rm eff} (u) \Bigr\rb\; .
\lab{EF-action}
\ee
We now observe an important result -- in \rf{EF-action} we have 
a dynamically created scalar field $u$ with a non-trivial effective scalar 
potential $U_{\rm eff}(u)$ \rf{U-eff} entirely {\em dynamically generated} 
by the initial non-Riemannian volume elements in \rf{NRVF-1} because of the
appearance of the free integration constants $M_1, M_2, \chi_2$ 
in their respective equations of motion \rf{A-eq}-\rf{C-B-eq}.

%%%%%%%%%%%%%%%%%%%%%%%%%%%%%%%%%%%%%%%%%%%%%%%%%%%%%%%%%%%%%%%%%%%%%%%
\begin{figure}[t!] %[H]
\centering
\includegraphics[width=0.49\textwidth]{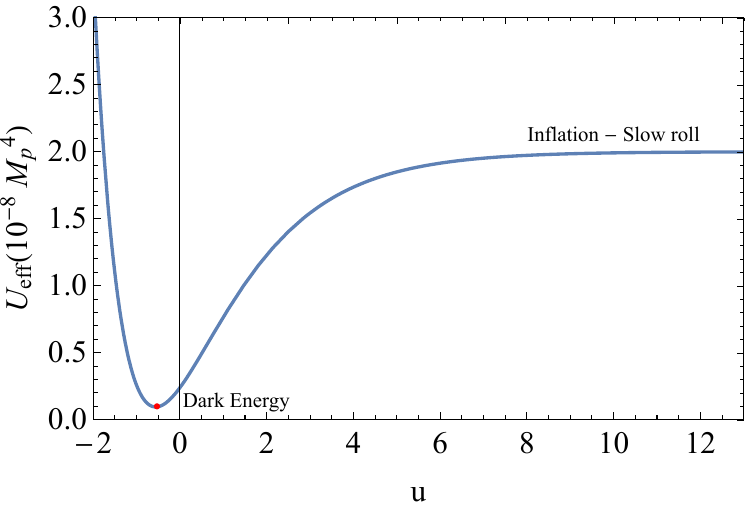}
\caption{Qualitative shape of the effective potential $U_{\rm eff} (u)$ in 
the Einstein frame, as presented in Eq. \rf{U-eff}. The physical unit for $u$ is
$M_{Planck}/\sqrt{2}$.}
% , for $\L_0 \sim M_1 \sim M_2 \sim 10^{-8} M_{Pl}^4$ and $\chi_2 \sim 1$.}
\label{fig1}
\end{figure}
%%%%%%%%%%%%%%%%%%%%%%%%%%%%%%%%%%%%%%%%%%%%%%%%%%%%%%%%%%%%%%%%%%%%%%%
The qualitative shape of \rf{U-eff} is depicted on Fig.1. The effective potential $U_{\rm eff} (u)$ has two main features relevant for
cosmological applications. First, $U_{\rm eff} (u)$ \rf{U-eff} possesses a flat region
for large positive $u$ and, second, it has a stable minimum for a small finite value $u=u_{*}$:

(i) $U_{\rm eff} (u) \simeq 2\L_0$ for large $u$;

(ii) $\partder{U_{\rm eff}}{u} = 0\;$ for $u\equiv u_{*}$ where:
\be
\exp\bigl(-\frac{u_{*}}{\sqrt{3}}\bigr) = \frac{M_1}{2\chi_2 M_2}  \quad ,\quad, 
\frac{\pa^2 U_{\rm eff}}{\pa u^2}\bgv_{u=u_{*}} = \frac{M_1^2}{6\chi_2 M_2} >0 \; . 
\lab{stable-min}
\ee
The flat region of $U_{\rm eff} (u)$ for large positive $u$ correspond to
``early'' universe' inflationary evolution with energy scale $2\L_0$. On the
other hand, the region around the stable minimum at $u=u_{*}$ \rf{stable-min}
correspond to ``late'' universe' evolution where the minimum value of the
potential:
\be
U_{\rm eff} (u_{*})= 2\L_0 - \frac{M_1^2}{4\chi_2 M_2} \equiv 2 \L_{\rm DE}
\lab{DE-value}
\ee
is the dark energy density value \cite{Angus:2018tko,Zhang:2018gbq}. 

Let us note that the effective potential $U_{\rm eff}$
\rf{U-eff} generalizes the well-known Starobinsky
inflationary potential \ct{Starobinsky:1979ty} (\rf{U-eff} reduces to Starobinsky
potential upon taking the following special values for the
parameters: $\L_0 = \frac{1}{4}M_1 = \h \chi_2 M_2$). 
% $U_{\rm eff}$).

%%%%%%%%%%%%%%%%%%%%%%%%%%%%%%%%%%%%%%%%%%%%%%%%%%%%%

\section{Evolution of the homogeneous solution}

We now consider reduction of the Einstein-frame action \rf{EF-action} to the
Friedmann-Lemaitre-Robertson-Walker (FLRW) setting with metric
$ds^2 = - N^2 dt^2 + a(t)^2 d{\vec x}^2$, and with $u=u(t)$. %and $\vp=\vp(t)$.
In order to study the evolution of the scalar field $u=u(t)$ and the Friedmann scale
factor $a=a(t)$, it is useful to use the method of 
autonomous dynamical systems. 

The FLRW action describes a minimally coupled canonical
scalar field $u$ with specific potential $U_{\rm eff}(u)$ \rf{U-eff} 
% plus a free massless scalar $\vp(t)$ 
(using again units with $16 \pi G_{\rm Newton} = 1$):
\br
S_{\rm FLRW} = \int d^4 x \Bigl\lb - 6\frac{a\adot^2}{N}
+ N a^3 \Bigl(% \h \frac{\vpdot^2}{N^2} +
\h \frac{\udot^2}{N^2} 
\nonu \\
+ M_1 e^{-u/\sqrt{3}} - M_2\chi_2 e^{-2u/\sqrt{3}} - 2\L_0\Bigr)\Bigr\rb \; .
\lab{EF-action-FLRW}
\er
Variations w.r.t. $N, a, u$ (and subsequently using the gauge $N=1$
for the lapse function) yield the pertinent Friedmann and field equations 
($H = \adot/a$ being the Hubble parameter):
\br
H^2 = \frac{1}{6}\rho \;\; ,\;\; \rho = % \h\vpdot^2 + 
\h\udot^2 + U_{\rm eff}(u)\; , 
\lab{Fried-1} \\
\Hdot = - \frac{1}{4} (\rho + p) 
\;\; ,\;\; p = % \h\vpdot^2 + 
\h\udot^2 - U_{\rm eff}(u)\; ,
% \Hdot = - 3H^2 +\h U_{\rm eff}(u)\Bigr) \; ,
\lab{Fried-2} \\
\uddot + 3H \udot + \partder{U_{\rm eff}}{u} = 0 \; .
\lab{u-eq}  % \\
% \vpddot + 3H \vpdot = 0 \;\; \to \;\; \vpdot=\frac{{\rm C_1}}{a^3} \; .
% \lab{vp-eq}
\er

In the treatment of Eqs.\rf{Fried-1}-\rf{u-eq} it is instructive to rewite 
them in terms of a set of dimensionless parameters (following the approach in
Ref.\ct{Bahamonde:2017ize}): 
\be
x := \frac{\dot{u}}{\sqrt{12} H},\quad 
y := \frac{\sqrt{U_{\rm eff}(u) - 2\L_{\rm DE}}}{\sqrt{6} H}, \quad 
z := \frac{\sqrt{\L_{\rm DE}}}{\sqrt{3}H} \; ,
\lab{xyz-def}
\ee
with $L_{\rm DE}$ as in \rf{DE-value}. 
In these coordinates the system defines a closed orbit:
\be
x^2 + y^2 + z^2 = 1 \; ,
% + \Omega_\text{stiff} = 1 \; ,
\lab{orbit}
\ee
which is equivalent to the first Friedmann equation \rf{Fried-1}.
% %%%%%%%%%%%%%%%%%%%%%%%%%%%%%%%%%%%%%%%%%%%%%%%%%%%%%%%%%%%%%
\begin{figure}[t!]
\centering
\includegraphics[width=0.49\textwidth,valign=t]{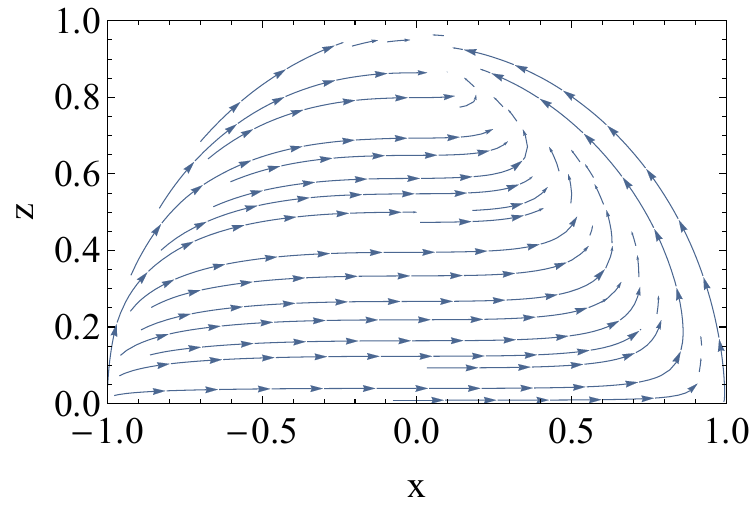}
\caption{Phase space portrait of the autonomous system \rf{asm}. The $x$ axis denotes the 
relative kinetic part of the scalar inflaton, and the $z$ axis denotes the relative part of 
the dark energy density $\L_{\rm DE}$.}
\label{fig2}
\end{figure}
Employing the variables $(x,y,z)$ in Eqs.\rf{Fried-1}-\rf{u-eq} and
taking into account the constraint \rf{orbit} we obtain
the autonomous dynamical system w.r.t. $(x,z)$:
\br
x' = \frac{\sqrt{3}}{2 \Lambda_{DE}} z^2 \left[ -M_1 \xi(x,z)
+2M_2 \chi_2 \xi^2(x,z)\right] 
-3x (1-x^2) \; , 
\nonu \\
% \lab{asm1} \\
z' = 3zx^2 \; , \phantom{aaaaaaaaaa}
\lab{asm}
\er
where the primes denote derivative w.r.t. the parameter $\cN = \log a$, and
the function $\xi(x,z)$ is defined as: 
\be
\xi(x,z) = \frac{M_1}{2\chi_2 M_2} \Bigl\lb 1 -
\sqrt{\frac{8\L_0 M_2\chi_2}{M_1^2}\,\frac{1-x^2 -z^2}{z^2}}\Bigr\rb \; .
\lab{zeta-def}
\ee

%%%%%% change-begin %%%%%%
The phase portrait of the system \rf{asm} is depicted on Fig.2. 
There are two critical points in the system. 
The stable point $A\(x=0, z=1 \)$ 
corresponds to the ``late'' universe de Sitter solution with the asymptotic 
cosmological constant $\L_{\rm DE}$ \rf{DE-value}.

The second point $B\(x=0,z=\sqrt{\L_{\rm DE}/\L_0}\)$ is unstable
corresponding to the beginning of the universe' evolution in the ``early'' universe at
large $u$. If the evolution starts at any point close to $B$, initially the
evolution is of de Sitter type with effective cosmological constant $\approx \L_0$.
Then the dynamics drives the system away from $B$ all the way towards the
stable point $A$ at late times.
%%%%%% change-end %%%%%%

\begin{figure*}[t!]
\centering
\includegraphics[width=0.49\textwidth,valign=t]{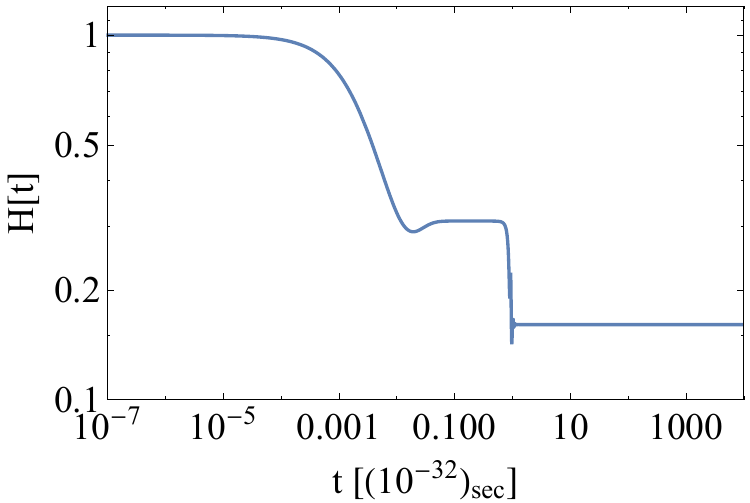}
\includegraphics[width=0.49\textwidth,valign=t]{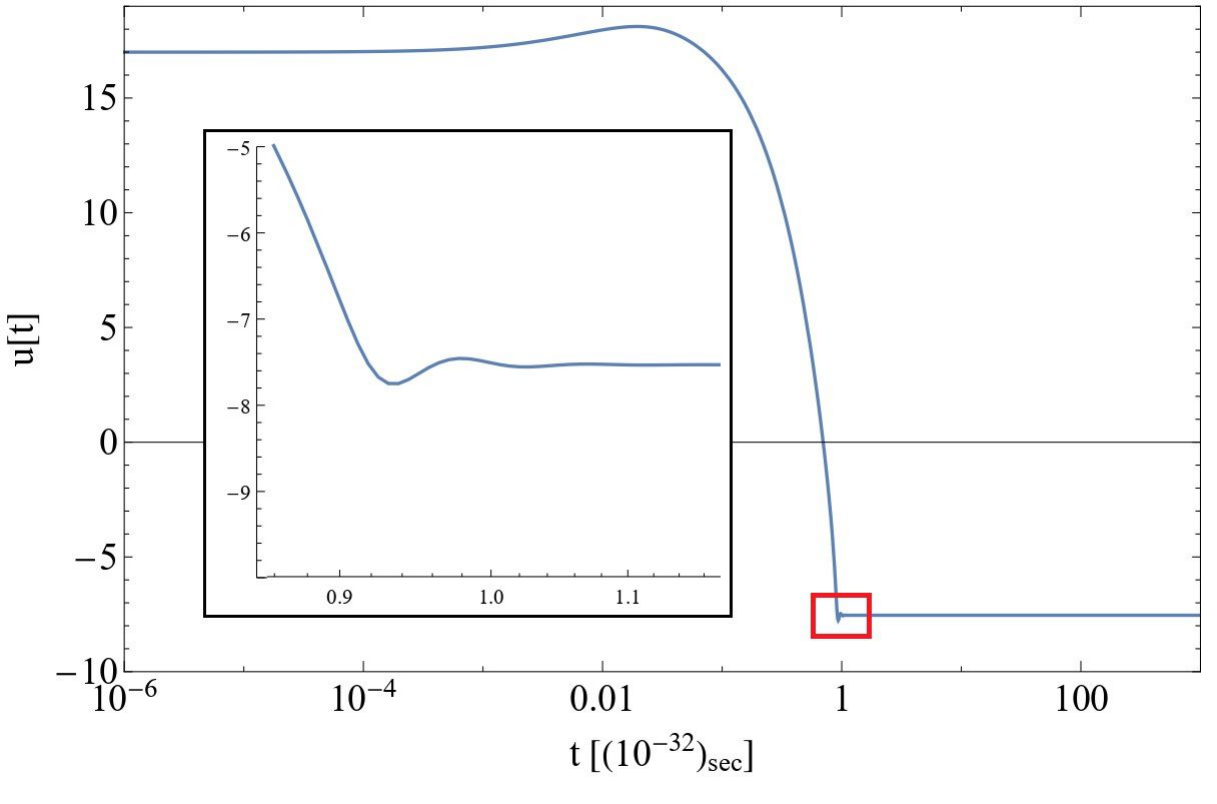}
\caption{Numerical example of the Hubble parameter $H(t)$ and the scalar field
$u(t)$ vs. time. 
For short times the inflationary Hubble parameter is large and afterwards 
approaches its cosmological late time value. As an example we take: 
$\frac{M_1}{M_2\chi_2} \simeq 10^{-2}$, $H(0) \sim 1$, $u(0) \sim 17$.
%%%%%%  change-begin %%%%%%
The physical units for the numbers representing $u$ and $H$ on the vertical axis 
in both graphics conform to our choice of normalization of ``Planck'' units 
$16\pi G_{Newton} = 1$ in Eq.\rf{NRVF-0} and henceforth.
Thus, the physical unit for $u$ is $1/\sqrt{16\pi G_{Newton}} \equiv M_{Planck}/\sqrt{2}$,
and the unit for $H$ is around $10^{-17} {\rm sec}^{-1}$ or around 
$300 {\rm (km/sec)}/{\rm Mpc}$ which conforms to the current value of the
Hubble parameter around $60 {\rm (km/sec)}/{\rm Mpc}$.
%%%%%%  change-end %%%%%%
On the right panel the
blown-up rectangle depicts the oscillations of $u(t)$ around the minimum of
$U_{\rm eff}$ \rf{U-eff}. One can see that the universe starts with an 
inflationary Hubble constant and ends with a smaller value  
representing the dark energy epoch.} 
\label{fig3}
\end{figure*}

Numerical solutions are demonstrated in Fig. \ref{fig3}. 
One can see that the Hubble parameter begins and ends with two different values. 
The first one is related to the inflationary epoch and the other related to the dark 
energy in the late universe. The scalar field $u$  oscillates around the minimum 
point $u_{*}$ \rf{stable-min} of $U_{\rm eff}$ \rf{U-eff}, 
which corresponds to particle creation in the reheating epoch. 

%%%%%%%%%%%%%%%%%%%%%%%%%%%%
%%%%%%%%%%%%%%%%%%%%%%%%%%%

\section{Perturbations}
In order to check the viability of the model we investigate the perturbations of 
the above background evolution, in particular focusing on the inflationary 
observables such as the scalar spectral index $n_s$ and the tensor-to-scalar 
ratio $r$. As usual, we introduce the Hubble slow-roll parameters \cite{Martin:2013tda,Bamba:2016wjm}, which in our case using the potential \rf{U-eff} read:
%%%%%% change-begin %%%%%%
\br
\epsilon = \Bigl(\frac{U_\text{eff}^\pr (u)}{U_\text{eff}(u)}\Bigr)^2
= \frac{4\z^2}{3} \frac{\bigl(1/2 - \z\bigr)^2}{\bigl\lb \bigl(1/2 - \z\bigr)^2
+ \d/4\bigr\rb^2} \; ,
\lab{eps-1} \\
|\eta| = 2 |\frac{U_\text{eff}^{\pr\pr}(u)}{U_\text{eff}(u)}|
= \frac{2\z}{3} \frac{\bigl(1-4\z\bigr)}{\bigl\lb \bigl(1/2 - \z\bigr)^2
+ \d/4\bigr\rb} \; ,
\lab{eta-1}
\er
where:
\be
\z\equiv \frac{M_2 \chi_2}{M_1}\,e^{-u/\sqrt{3}} \quad,\quad
\d \equiv \frac{8M_2 \chi_2}{M_1^2} \L_{\rm DE} \; ,
\lab{zeta-delta-def}
\ee
with $\L_{\rm DE}$ -- the dark energy density \rf{DE-value}, and therefore $\d$ very
small.

Inflation ends when $\epsilon (u_f) = 1$ for some $u=u_f$ where
($\z_f \equiv \frac{M_2 \chi_2}{M_1} e^{-u_f/\sqrt{3}}$):
\br
\z_f =
\frac{1}{2\bigl(1+2/\sqrt{3}\bigr)} \Bigl\lb 1+\frac{1}{\sqrt{3}}
- \sqrt{1/3 - \bigl(1+2/\sqrt{3}\bigr)\d} \Bigr\rb 
\nonu \\
\simeq \frac{1}{2\bigl(1+2/\sqrt{3}\bigr)} \; . \phantom{aaaaaaaa}
\lab{z-f}
\er

For the number of $e$-foldings 
$\cN = \h \int_{u_i}^{u_f} du \; U_{\rm eff}/ U_{\rm eff}^\pr$ we obtain:
\br
\cN = \frac{3}{8}(1+\d)\Bigl(1/\z_i - 1/\z_f\Bigr) 
\nonu \\
- \frac{3}{4}(1-\d) \log\frac{\z_f}{\z_i}
+ \frac{3}{4}\d\,\log\Bigl(\frac{1-2\z_i}{1-2\z_f}\Bigr) \; ,
\lab{N-def}
\er
where $\z_i \equiv \frac{M_2 \chi_2}{M_1} e^{-u_i/\sqrt{3}}$ and $u=u_i$ is
very large corresponding to the start of the inflation. Ignoring $\d$ and
using the last equality \rf{z-f} we have approximately:
\br
\cN \simeq \frac{3M_1}{8M_2\chi_2} e^{u_i/\sqrt{3}} - \frac{\sqrt{3}}{4} u_i
- \frac{3}{4} \bigl(1+2/\sqrt{3}\bigr) 
\nonu \\
+ \frac{3}{4} \log\Bigl(2\bigl(1+2/\sqrt{3}\bigr)\Bigr) \; .
\lab{N-approx}
\er
%%%%%% change-end %%%%%%

Using the slow-roll parameters, one can calculate the values of the scalar 
spectral index   and the tensor-to-scalar ratio respectively as  
\cite{Nojiri:2019kkp,Dalianis:2018frf}:
\begin{equation}
r \approx 16 \epsilon, \quad n_s \approx 1- 6\epsilon + 2 \eta 
\end{equation}
Taking into account Eqs.\rf{eps-1},\rf{eta-1} (ignoring $\d$) and \rf{N-approx} 
we find:
\br
r \simeq \frac{12}{\Bigl\lb \cN + \frac{\sqrt{3}}{4} u_i(\cN) + c_0\Bigr\rb^2} 
\; ,
\lab{ns-r-approx} \\
c_0 \equiv \frac{\sqrt{3}}{2} - \frac{3}{4} \log\Bigl(2\bigl(1+2/\sqrt{3}\bigr)\Bigr) \; ;
\nonu
\er
and 
\begin{equation}
n_s \simeq 1 -\frac{r}{4}-\sqrt{\frac{r}{3}},
\end{equation}
where $u_i (\cN)$ is the solution of the transcedental Eq.\rf{N-approx} for $u_i$ as a function of $\cN$.

One viable example in our model is to take $\cN = 60$ $e$-folds. 
%%%%%% change-begin %%%%%%
Eq.\rf{N-approx} yields $\cN = 60$ provided we choose 
$\frac{M_1}{M_2\chi_2} \simeq 10^{-2}$, which yields $u_i \simeq 17$. 
%%%%%% change-end %%%%%%
In such a way the observables are predicted to be:
\begin{equation}
n_s \approx 0.969, \quad r \approx 0.0026,
\end{equation}
which are well inside the PLANCK observed constraints \cite{Akrami:2018odb}:
\begin{equation}
0.95 < n_s < 0.97, \quad r < 0.064
\end{equation}

\begin{figure}[t!]
\centering
\includegraphics[width=0.49\textwidth]{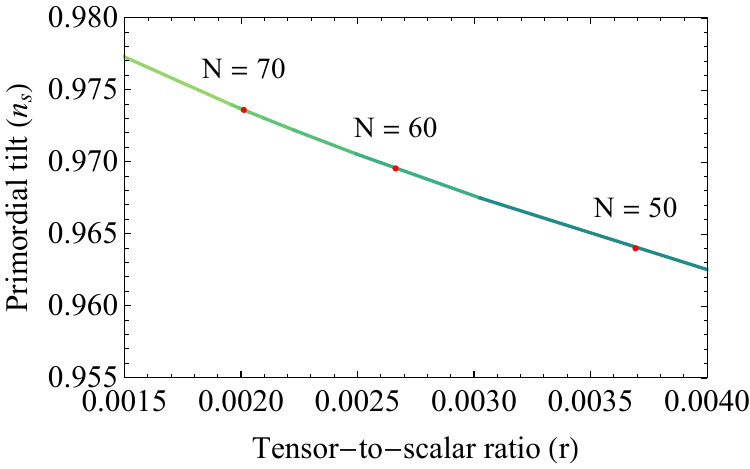}
\caption{The predicted values of the $r$ and $n_s$ for different $e$-foldings. The different values of the $r$ and $n_s $ are compatible with the observational data.}
\label{fig4}
\end{figure}

Fig. \ref{fig4} demonstrates the relation between the number of $e$-folds 
and the dimensionless parameters. One can see that all those values 
fit the latest PLANCK collaboration constraints.
%%%%%%%%%%%%%%%%%%%%%%%%%%%%%%%%%%%%%%%%%%

\section{Conclusions}
We propose a very simple gravity model without any initial matter fields 
in terms of several alternative 
non-Riemannian spacetime volume elements within the second order (metric) 
formalism. We show how the non-Riemannian volume-elements,
when passing to the physical Einstein frame, create a canonical scalar
field and produce dynamically a non-trivial inflationary-type potential for
the latter with a large flat region and a stable low-lying minimum.
We study the evolution of the cosmological solutions from the point of view
of the theory of dynamical systems. Our model predicts scalar spectral index 
$n_s \approx 0.96$ and tensor-to-scalar ratio $r \approx 0.002$ for 60 $e$-folds, 
which is in accordance with the observational data.

A natural next step is to consider two-field inflation
(Refs. \ct{Bahamonde:2017ize,2-field-inflation-1,2-field-inflation-1a,2-field-inflation-2, Ben-Dayan:2018mhe,Artymowski:2019vfy,Ben-Dayan:2014isa,Benisty:2019bmi,Aragam:2019khr,Kubo:2018kho}, for a
geometric treatment see Refs. \ct{lili,MadrizAguilar:2019wgd,Barvinsky:2019qzx,Ahmad:2019jbm,Granda:2019jqy}, and references therein) by adding a new scalar field $\vp$ with non-trivial potentials
in the starting modified gravity action \rf{NRVF-1} built 
in terms of several non-Riemannian volume elements and subject to preserving the requirement of global Weyl-scale invariance \rf{scale-transf}.In this case the non-Riemannian volume elements will again generate a second scalar field $u$ and create dynamically a non-trivial two-field scalar potential with a very specific geometry of the field space of $\vp, u$.
%% ADDED IN PROOF
This is studied in more detail in our subsequent work \cite{Benisty:2019bmi}, where it is shown that the latter dynamically generated two-field inflationary model similarly conforms to the observational data.

\begin{acknowledgements}
We gratefully acknowledge support of our collaboration through the Exchange
Agreement between Ben-Gurion University, Beer-Sheva, Israel and Bulgarian
Academy of Sciences, Sofia, Bulgaria. E.N. and S.P. are thankful for support
by Contract DN 18/1 from Bulgarian National Science Fund. D.B., E.G. and
E.N. are also partially supported by COST Actions CA15117, CA16104 and the action CA18108.
D.B., E.N. and S.P. acknowledge illuminating discussions with Lilia Anguelova.
%%%%%% change-begin %%%%%%
Finally we would like to thank the referees whose comments contributed to
significant improvement of the presentation.
%%%%%% change-end %%%%%%
\end{acknowledgements}

%\bibliographystyle{unsrt}
%\bibliography{ref}   % name your BibTeX data base


\end{document}